\documentclass[12pt, preprint]{aastex}
\begin{document}
\title{TURNING AROUND ALONG THE COSMIC WEB}
\author{Jounghun Lee\altaffilmark{1} and 
Gustavo Yepes\altaffilmark{2,3}}
\altaffiltext{1}{Astronomy Program, Department of Physics and Astronomy, Seoul National University, 
Seoul 08826, Korea
\email{jounghun@astro.snu.ac.kr}}
\altaffiltext{2}{Departamento de F\'{\i}sica Te\'orica, Universidad Aut\'onoma de Madrid, 28049 Madrid, Spain}
\altaffiltext{3}{ASTRO UAM, UAM Unidad Asociada CSIC.}
\begin{abstract}
A bound-violation designates a case that the turn-around radius of a bound object exceeds the upper limit put by the 
spherical collapse model based on the standard $\Lambda$CDM paradigm. Given that the turn-around radius 
of a bound object is a stochastic quantity and that the spherical model overly simplifies the true gravitational collapse which 
actually proceeds anisotropically along the cosmic web,  the rarity of the occurrence of a bound violation may depend 
on the web environment. 
Assuming a Planck cosmology, we numerically construct the bound-zone peculiar velocity profiles along the cosmic web 
(filaments and sheets) around the isolated groups with virial mass $M_{\rm v}\ge 3\times 10^{13}\,h^{-1}M_{\odot}$ identified in the 
Small MultiDark Planck simulations and determine the radial distances at which their peculiar velocities equal the Hubble expansion 
speed as the turn-around radii of the groups.  It is found that although the average turn-around radii of the isolated groups are well 
below the spherical bound-limit on all mass scales, the bound violations are not forbidden for individual groups and that the cosmic 
web has an effect of reducing the rarity of the occurrence of a bound violation.   Explaining that the spherical bound 
limit on the turn-around radius in fact represents the threshold distance up to which the intervention of the external 
gravitational field in the bound-zone peculiar velocity profiles around the {\it non-isolated} groups stays negligible, 
we discuss the possibility of using the threshold distance scale to constrain locally the equation of state of dark energy .
\end{abstract}
\section{INTRODUCTION}
\label{sec:intro}

In a $\Lambda$CDM (the cosmological constant $\Lambda$ + cold dark matter) universe, a mildly overdense site which accretes 
masses from the surrounding with growing comoving radius will never encounter the turn-around moment since its 
gravity will be incapable of winning over the accelerating expansion of spacetime driven by such inert dark energy as $\Lambda$. 
Therefore, only those overdense sites that have already formed bound objects until the present epoch or at least turned 
around before the onset of the accelerating phase of spacetime can possess finite turn-around radii, defined as the effective 
radii of the sites at the moment when their gravitational force begins to hold down the Hubble expansion \citep{PT14,pavlidou-etal14}.  
In other words,  the standard $\Lambda$CDM cosmology generically predicts the existence of a finite upper limit on the 
turn-around radii of the bound objects observed in the local universe. 

Pointing out that the turn-around radius of a bound object, $r_{\rm t}$, is a well defined linear quantity, \citet{PT14} found 
an analytic expression for the upper limit , $r_{\rm t, u}$, on the possible values of $r_{\rm t}$ under one simplified 
assumption that the gravitational collapse after the turn-around moment proceeds isotropically: 
\begin{equation}
\label{eqn:rt_u}
r_{\rm t, u} = \left(\frac{ 3MG}{\Lambda c^{2}}\right)^{1/3}\, ,
\end{equation}
Strictly speaking, $M$ in the RHS of Equation (\ref{eqn:rt_u}) denotes the mass enclosed by the turn-around radius $r_{\rm t}$ 
the ratio of which to virial mass $M_{\rm v}$ depends on mass scales and redshifts. Notwithstanding, \citet{PT14} suggested that  
the upper limit of the turn-around radius be expressed in terms of a readily measurable mass like the virial mass since the 
mass enclosed by $r_{\rm t}$ cannot be independently measured in practice. See \citet{PT14} for the detailed 
explanations.  Throughout this paper, we approximate $M$ in the RHS of Equation (\ref{eqn:rt_u}) by $M_{\rm v}$.

Equation (\ref{eqn:rt_u}) predicts that the average turn-around radius on the mass scale $M_{\rm v}$ stays much lower than 
$r_{\rm t, u}$. If a substantially large number of the objects in the local universe should be found to exhibit a bound violation having 
larger turn-around radii than $r_{\rm t, u}$ given in Equation (\ref{eqn:rt_u}), then it would challenge the $\Lambda$CDM model.  
As mentioned in \citet{PT14}, the upside of using the turn-around radii of the bound objects as a test of the $\Lambda$CDM 
model lies in the fact that it requires only local observables whose systematics would be better controlled unlike 
the standard global diagnostics which usually require to measure the growth and/or expansion history of the whole universe 
\citep[see][and references therein]{linder03,linder05}.

In the conventional approach, the values of $r_{\rm t}$ of the galaxy groups and clusters observed in the local universe 
have been estimated by directly measuring the distances to the surfaces of the zero radial velocities of the bound-zone galaxies 
\citep[e.g., see][and references therein]{kara-etal14}. However, the estimates of $r_{\rm t}$ based on this 
conventional method has been known to suffer from low accuracy that is imputable to notoriously large uncertainties in the 
measurements of the peculiar velocities of the bound-zone galaxies around the groups and clusters \citep[see][]{PT14}.  
Very recently, \citet{lee-etal15} proposed a new methodology to overcome the low accuracy in the estimation of $r_{\rm t}$ and 
applied it to a nearby galaxy group NGC 5353/4 \citep{TT08,tully15}.  Their methodology turned largely to the numerical claim 
of \citet{falco-etal14} that the peculiar velocity profile in the bound zone abound a massive group takes on a universal form and that 
the profile can be readily reconstructed from the two-dimensional spatial distributions of the bound-zone galaxies along the cosmic 
web. 

The turn-around radius of NGC5353/4 estimated by \citet{lee-etal15} with their new method was found to commit a bound violation.
That is, the value of $r_{\rm t}$ of the NGC 5353/4 group has been found to exceed the upper limit $r_{\rm t, u}$ given in 
Equation (\ref{eqn:rt_u}). Being cautious and conservative, \citet{lee-etal15} have suggested that the robustness of the new 
method as well as the validity of the universal formula for the peculiar velocity profile proposed by \citet{falco-etal14}  should be 
thoroughly examined before putting a proper interpretation of their result.
As clearly mentioned in \citet{PT14},  the turn-around radius of an object is a stochastic quantity, varying in a broad range. 
Equation (\ref{eqn:rt_u}) puts the upper  limit on the {\it average} turn-around radius on the mass scale $M_{\rm v}$ but not 
on all turn-around radii of individual objects with virial mass $M_{\rm v}$.  In other words, it might not be forbidden even in the 
$\Lambda$CDM cosmology to observe an individual object with mass $M_{\rm v}$ whose turn-around radius exceeds the 
spherical bound-limit. What Equation (\ref{eqn:rt_u}) truly implies is that it would be quite rare to witness a bound violation 
in the $\Lambda$CDM universe.
 
Besides, what has not been properly taken into account in the analytic derivation of $r_{\rm t, u}$ by \citet{PT14} is the 
fact that the true gravitational collapse proceeds quite anisotropically \citep{BM96} which is responsible for the eventual 
formation of the cosmic web in the universe \citep{web96}.  Although \citet{lee-etal15} used the same analytic formula 
proposed by \citet{falco-etal14} to make an estimate of the turn-around radius of NGC 5353/4 along the filament, it has yet 
to be tested whether or not the peculiar velocity profile constructed along the cosmic web would still be validly approximated by 
the same formula that was obtained by taking the isotropic average over all directions. 
Thus, the remaining critical questions are (1) how rare occasion a bound violation is in a $\Lambda$CDM universe; and 
(2) how the true gravitational collapse that tends to proceed anisotropically along the cosmic web affects the rarity of the occurrence 
of bound violation, both of which we attemp to answer in this work.

The contents of this Paper can be summarized as follows. In Section \ref{sec:review} a brief review of the methodology of 
\citet{lee-etal15} and its key assumptions are presented. In Section \ref{sec:web} a comprehensive numerical test of the methodology 
is presented and the effect of the web environment on the construction of the peculiar velocity profiles is explored. 
In Section \ref{sec:iso} the estimates of the turn-around radii along the cosmic web on various mass scales are presented. 
In Section \ref{sec:con} the final results are summarized and the physical implications are discussed. 

\section{METHODOLOGY AND THE KEY ASSUMPTIONS: A REVIEW}\label{sec:review}
 
Let us consider a spatial point at a distance, $r$, from the center of a massive group with virial radius, $r_{\rm v}$,  in a space 
expanding at a global rate, $H_{\rm 0}$. From here on, the radial component of the peculiar velocity field in the direction from the 
halo center to the spatial point is called {\it the peculiar velocity at $r$} for short. 
If the distance $r$ is included in the bound zone range of $3\le r/r_{\rm v}\le 8$ where the subspace expands at a lower rate than 
$H_{\rm 0}$ due to the net gravitational effect of the main group, the spherically averaged value of of the peculiar velocity, 
$v_{\rm p}$, at $r$ is well depicted by the following universal formula with two parameters $a$ and $b$ \citep{falco-etal14}: 
\begin{equation}
\label{eqn:vpr}
v_{p}(r) = - a\,V_{\rm c}\left(\frac{r}{r_{\rm v}}\right)^{-b}\, ,
\end{equation}
where $V_{\rm c}$ is the magnitude of the circular velocity of the main group at its virial radius given as $V^{2}_{c}=GM_{\rm v}/r_{\rm v}$ 
with virial mass $M_{\rm v}$.  Fitting Equation (\ref{eqn:vpr}) to the numerical results from a high-resolution $N$-body simulation, 
\citet{falco-etal14} found $a=0.8\pm 0.2$ and $b=0.42\pm 0.16$ and speculated that these best-fit values of the two parameters 
would be insensitive to the redshifts as well as to the mass scales.

Inspired by the existence of the universal peculiar velocity profile in the bound zone, \citet{lee-etal15} suggested that the 
turn-around radius, $r_{\rm t}$, of a main group be readily obtained just as the value at which the following relation is satisfied: 
\begin{equation}
\label{eqn:rt}
H_{\rm 0}r_{\rm t} = a\,V_{\rm c}\left(\frac{r_{\rm t}}{r_{\rm v}}\right)^{-b}\, ,
\end{equation}
where the left-hand side (LHS) represents the global expansion speed at $r_{\rm t}$ and the right-hand side (RHS) is the peculiar 
velocity at the same radius predicted by Equation (\ref{eqn:vpr}). Provided that the virial radius (or equivalently, virial mass) of a 
massive group is already known as a prior and that the two parameters in the RHS, $a$ and $b$, have truely constant values of 
$a\sim 0.8$ and $b\sim 0.42$ as speculated by \citet{falco-etal14}, solving Equation (\ref{eqn:rt}) for $r_{\rm t}$ should allow one 
to estimate the turn-around radius of a given group without necessitating an accurate measurement of the peculiar velocity field in 
its bound zone.

Although \citet{falco-etal14} showed that Equation (\ref{eqn:vpr}) with the constant values of $a$ and $b$ matched well 
the numerically obtained peculiar velocity profiles of three different cluster-size halos, it is inconclusive whether or not the 
bound-zone peculiar velocity profile is truly universal since the variations of its shape with mass and redshift were reported by other 
numerical works \citep[e.g.,][]{cuesta-etal08,lee16}.  To account for the possibility that the bound-zone peculiar velocity profile may 
not be universal, it is unavoidable to regard $a$ and $b$ in Equation (\ref{eqn:vpr}) as two {\it varying parameters} and to determine 
their best-fit values by fitting Equation (\ref{eqn:vpr}) to the bound-zone peculiar velocity profile constructed from observational data 
before estimating $r_{\rm t}$. 

The practical feasibility of constructing the bound-zone peculiar velocity profiles was improved by \citet{falco-etal14} who showed 
that if the average of the peculiar velocity field was taken over the spatial points distributed along the cosmic web (filaments and 
sheets), $v_{\rm p}(r)$ can be expressed in terms of directly measurable quantities: 
\begin{equation}
\label{eqn:2dvpr}
\frac{c\Delta z}{\cos\beta} - H_{0}\frac{R}{\sin\beta} = 
- aV_{\rm v}\left(\frac{R}{\sin\beta\,r_{\rm v}} \right)^{-b}\, . 
\end{equation}
Here $\Delta z$ represents the difference in redshifts between the main group and a bound-zone object, $R$ is the projected 
distance of the bound-zone object from the group center in the plane of sky, and $\beta$ is the inclination angle between the 
direction of the cosmic web and the line-of-sight, which are all directly measurable without making any specific assumption about 
the background cosmology.  \citet{lee-etal15} determined the values of $a$ and $b$ via adjusting the RHS to the LHS in Equation 
(\ref{eqn:2dvpr}) that can be also directly computed from the observables.  Applying this method to the bound zone of the NGC 
5353/4 group around which a narrow filamentary structure was detected (S.Kim in private communication), \citet{lee-etal15} 
pulled it off to estimate its turn-around radius.

\section{EFFECT OF THE COSMIC WEB ON THE VELOCITY PROFILE}\label{sec:web}

The halo catalog from the Small MultiDark Planck simulation (SMDPL) at $z=0$ is utilized to perform our numerical investigation. 
As a part of MultiDark and Bolshoi Project \citep{mdark_sim}, SMDPL was conducted with $3840^{3}$ particles of individual 
mass $9.63\times 10^{7}\,h^{-1}M_{\odot}$ in a periodic box of a volume $400^{3}\,h^{-3}\,{\rm Mpc}^{3}$ \citep{mdark_sim} 
for a Planck cosmology, i.e., a flat $\Lambda$CDM cosmology with $h=0.68,\ \Omega_{\Lambda}=0.69,\ 
\Omega_{\rm m}=0.31,\ \Omega_{\rm b}=0.048,\ \sigma_{8}=0.82,\ n_{s}=0.96$ \citep{planck14}. 
The SMDPL halo catalog at $z=0$ provides information on the comoving position (${\bf x}$), peculiar velocity (${\bf v}$), 
virial radius and mass ($r_{\rm v}$ and $M_{\rm v}$) of each dark matter halo identified by applying the Rockstar halo finder 
to the phase space distributions of the particles  \citep{rockstar_er}.  

The massive Rockstar halos with virial mass $M_{\rm v}\ge 3\times 10^{13}\,h^{-1}\,M_{\odot}$ are selected from the SMDPL 
catalog as the {\it main groups} whose bound-zone peculiar velocity profile and turn-around radii are to be estimated in the 
current work.  The lower-mass halos $M_{\rm v}< 3\times 10^{13}\,h^{-1}\,M_{\odot}$ are excluded from the sample of the main 
groups since the number of the neighbor halos in their bound zones is not large enough to construct the peculiar velocity profiles. 
A total of $8476$ halos in the SMDPL catalog at $z=0$ are found to be above the imposed mass cut-off and thus included in our 
sample of the main groups. Throughout this analysis we focus only on the present epoch ($z=0$) to match the real observations 
since the prior information in the virial radius and mass of the main groups required for our analysis is available with high accuracy 
only for the case of the objects located in the nearby universe. 

The region inside a spherical shell surrounding each main group with inner (outer) radius of $3r_{\rm v}$ ($8r_{\rm v}$) from the 
group center is designated as its bound zone. To construct the peculiar velocity profile from the halos located in the bound zone 
around each main group, we consider only those well resolved halos composed of $500$ or more particles, which should help 
us avoid numerical flukes. Basically, we search for those halos with $M_{\rm v}\ge 5\times 10^{10}\,h^{-1}\,M_{\odot}$ 
in the bound zone of each main group and then measure their separation distance vectors, ${\bf r}$, 
from the group center. The relative peculiar velocity of each bound-zone halo, $v_{\rm p}$ is determined as 
$v_{\rm p}=\left({\bf v}_{h}-{\bf v}_{g})\cdot{\bf r}\right/\vert{\bf r}\vert$ 
where ${\bf v}_{h}$  and ${\bf v}_{g}$ are the peculiar velocities of a bound-zone halo and a main group, respectively.  

To construct the peculiar velocity profile from the bound zone halos distributed along all directions, we first rescale the separation 
distance and the peculiar velocity of each bound-zone halo as $\tilde{r}=r/r_{\rm v}$ and $\tilde{v}_{\rm p}=v_{\rm p}/V_{\rm c}$, 
respectively,  where $r_{\rm v}$ and $V_{\rm c}$ are the virial radius and circular velocity of its main group, respectively. 
Dividing the range of $\tilde{r}$ into ten short intervals, we calculate the mean value of the rescaled peculiar velocity at each 
$\tilde{r}$-interval. Taking the ensemble average of the mean peculiar velocity at each $\tilde{r}$-interval over all of 
the selected main groups, we numerically construct $\tilde{v}_{p}(\tilde{r})$. 
The errors associated with $\tilde{v}_{p}(\tilde{r})$ is also calculated as one standard deviation in the measurement of the 
ensemble average at each $\tilde{r}$-interval. We finally fit the resulting numerical data points, $\tilde{v}_{\rm p}(\tilde{r})$, 
to Equation (\ref{eqn:vpr}) by adjusting the values of the two parameters, $a$ and $b$, with the help of the 
maximum-likelihood method \citep{WJ12}. The best-fit values are determined to be $a=0.74\pm 0.01$ and $b=0.31\pm 0.01$ 
where the marginalized error in the determination of each parameter is obtained by treating the other parameter as a 
nuisance one. 
 
\begin{figure}[tb]
\includegraphics[scale=1.0]{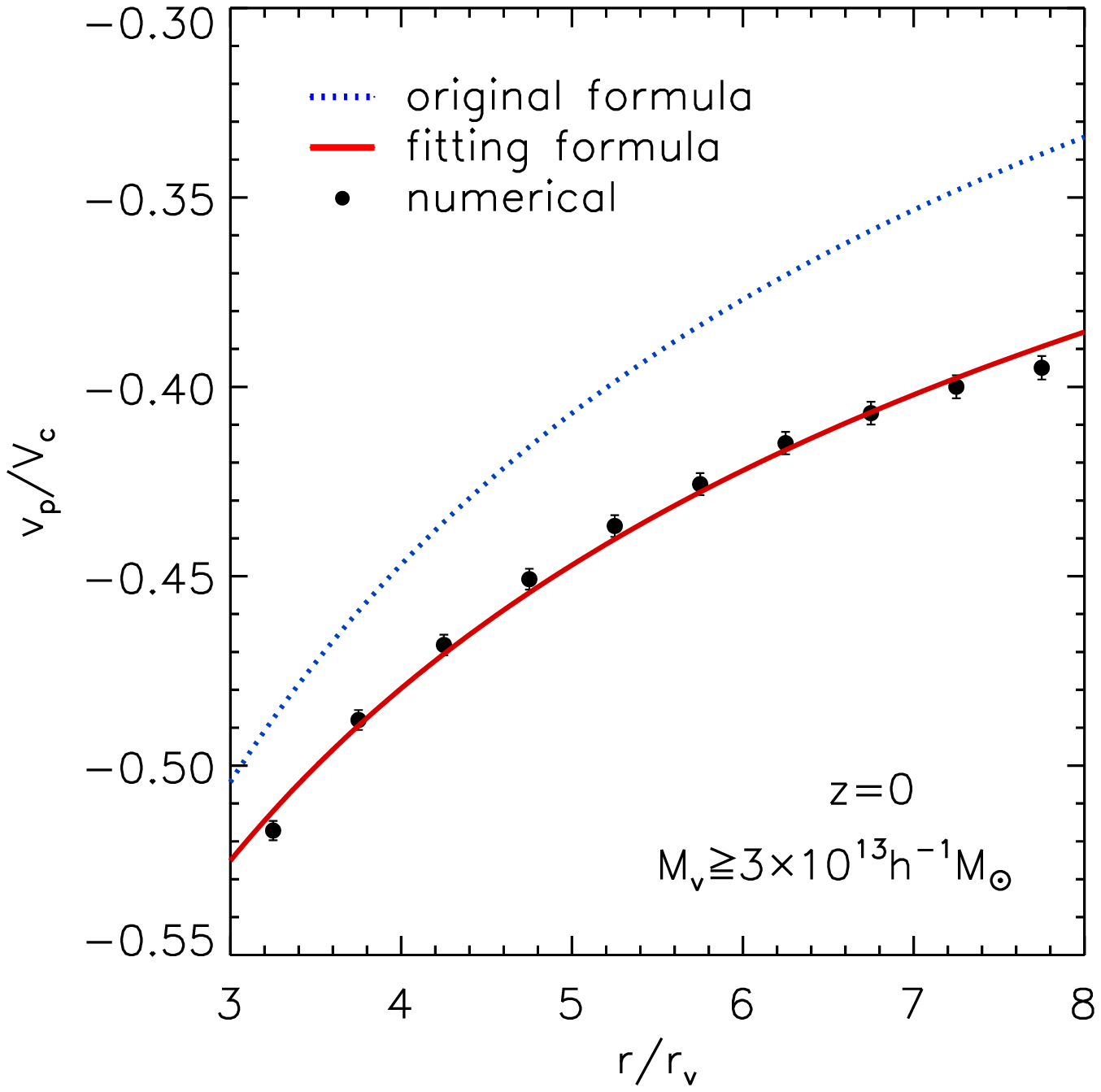}
\caption{Average peculiar velocity profile of the dark matter halos located in the bound zone around 
the massive central halos with virial mass $M_{\rm v}\ge 3\times 10^{13}\,h^{-1}\,M_{\odot}$ at $z=0$. 
The black filled circles with errors represent the numerical results from the Multi-Dark simulations 
and the red solid line represents the analytic formula with the best-fit parameters. The blue dotted 
lines corresponds to the original formula derived by \citet{falco-etal14}.}
\label{fig:vpr_z0}
\end{figure}

In Figure \ref{fig:vpr_z0}, the black filled circles represent the numerical result of $\tilde{v}_{\rm p}$ with errors while the red solid 
line displays the analytic formula, Equation (\ref{eqn:vpr}), with the best-fit values $a=0.74$ and $b=0.31$. The blue dotted line is 
the analytic formula with the two parameters set at the original values of $a=0.8$ and $b=0.42$ given by \citet{falco-etal14}.
Figure \ref{fig:vpr_z0} clearly reveals that although the best-fit values of $a$ and $b$ are determined to be different from the 
original values given in \citet{falco-etal14}, Equation (\ref{eqn:vpr}) with the best-fit parameters successfully describes the 
behavior of the bound-zone peculiar velocity profile, which is consistent with the result of \citet{lee16}. 

\begin{figure}[tb]
\includegraphics[scale=0.8]{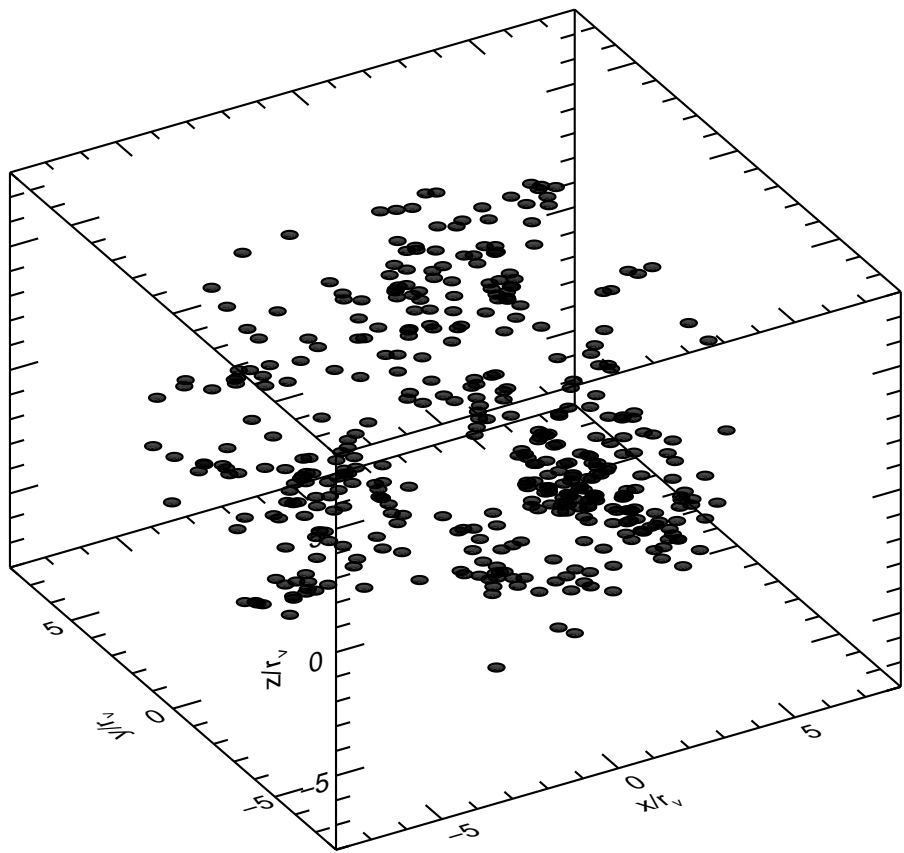}
\caption{Example of three dimensional spatial distributions of the bound-zone halos around a 
massive centeral halo resolved in the Multi-dark simulation at $z=0$.}
\label{fig:slice3d}
\end{figure}

Now, we want to explore how the best-fit values of $a$ and $b$ change when the construction of the mean peculiar velocity 
profile in the bound zone is done along the cosmic web.  Three dimensional spatial distributions of the bound-zone halos around 
a main group randomly selected from our sample is illustrated in Figure \ref{fig:slice3d}, which gives a glimpse of a specific 
direction along which the number density of the bound-zone halos is conspicuously high, corresponding to the direction of a 
bound-zone filament. As expected, the bound-zone halos are not distributed isotropically.

\begin{figure}[tb]
\includegraphics[scale=1.0]{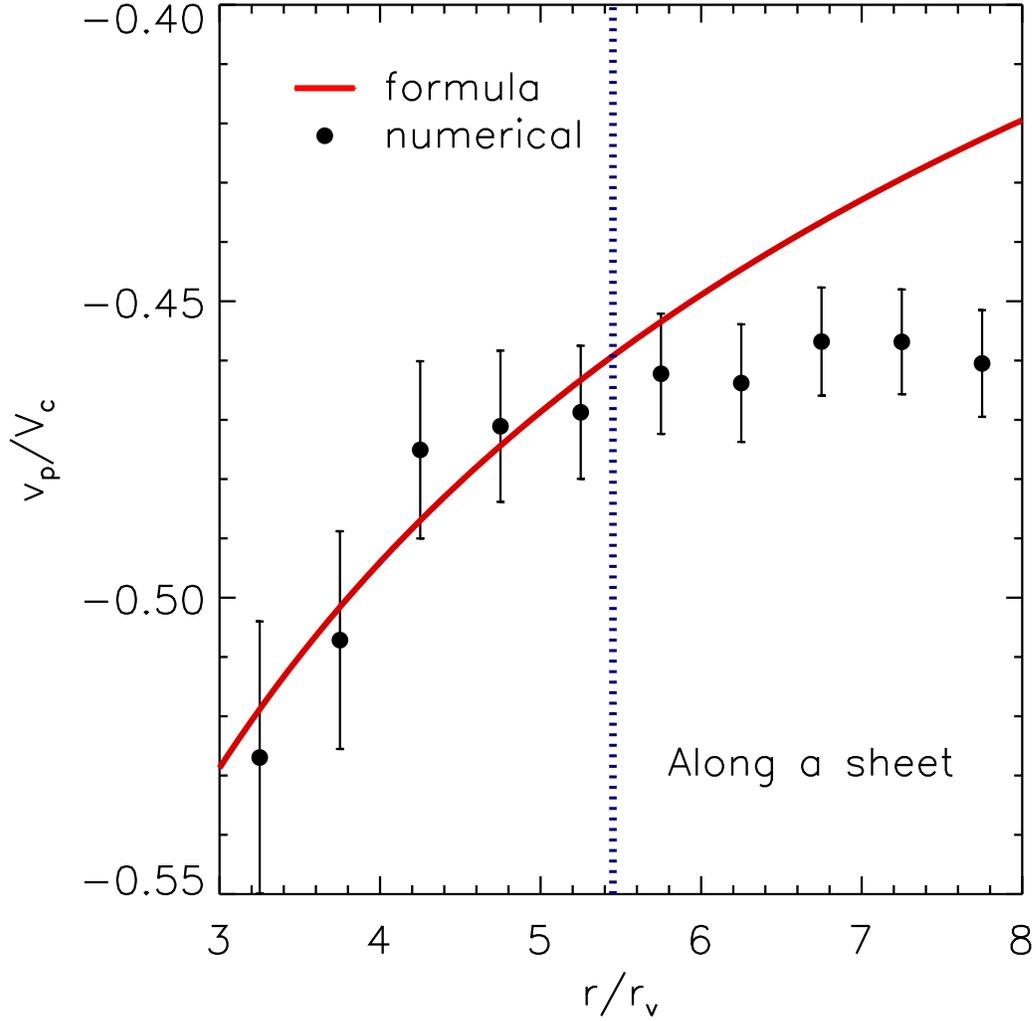}
\caption{Average peculiar velocity profile constructed along the bound-zone sheet. The black filled 
circles with errors represent the numerical results from the Small MultiDark Planck simulations \citep{mdark_sim} 
and the red solid line represents the fitting formula. The blue dotted line corresponds to the constant ratio of the 
bound-limit to the virial radius that is predicted to be a constant value in the Planck cosmology.}
\label{fig:vpr_sheet}
\end{figure}

To construct the peculiar velocity profile of each main group along the bound-zone sheet, we first  measure the cosines of the 
polar angles of the position vectors, ${\bf r}$, of the bound-zone objects from the center of each main group. 
Then, we divide the volume into ten slices through binning the values of $\cos\theta$ and count the number of the bound-zone 
objects whose values of $\cos\theta$ belong to each bin.  
Locating the slice to which the largest number of the bound-zone objects belong and designating it as the bound-zone sheet, 
we include only those objects belonging to the bound-zone sheet to calculate the mean rescaled peculiar velocity profile for 
each main group. Taking the ensemble average of the profiles over all of the main groups, we finally construct the peculiar 
velocity profile from the bound-zone halos distributed along the bound-zone sheet, which is shown as the black filled circles 
in Figure \ref{fig:vpr_sheet}. 

As can be seen, in the large $\tilde{r}$-section ($\tilde{r}> 5.5$), the magnitude of $\tilde{v}_{\rm p}(\tilde{r})$ does not 
monotonically decrease with $\tilde{r}$ unlike the case that the construction of $\tilde{v}_{\rm p}(\tilde{r})$ is performed 
from the bound-zone halos distributed along all directions (see Figure \ref{fig:vpr_z0}).  
Rather, it has a minimum value at $\tilde{r}\sim 5.5$ and becomes almost constant in the range of $\tilde{r}\ge 5.5$. 
Fitting Equation (\ref{eqn:vpr}) only to those numerical data points in the range of $\tilde{r}\le 5.5$ yields $a=0.69\pm 0.01$ 
and $b=0.24\pm 0.01$.  The red solid line in Figure \ref{fig:vpr_sheet} corresponds to the analytic model with 
these best-fit parameters. In the restricted range of $\tilde{r}\le 5.5$, the best-fit formula still works well, as can be seen.

\begin{figure}[tb]
\includegraphics[scale=1.0]{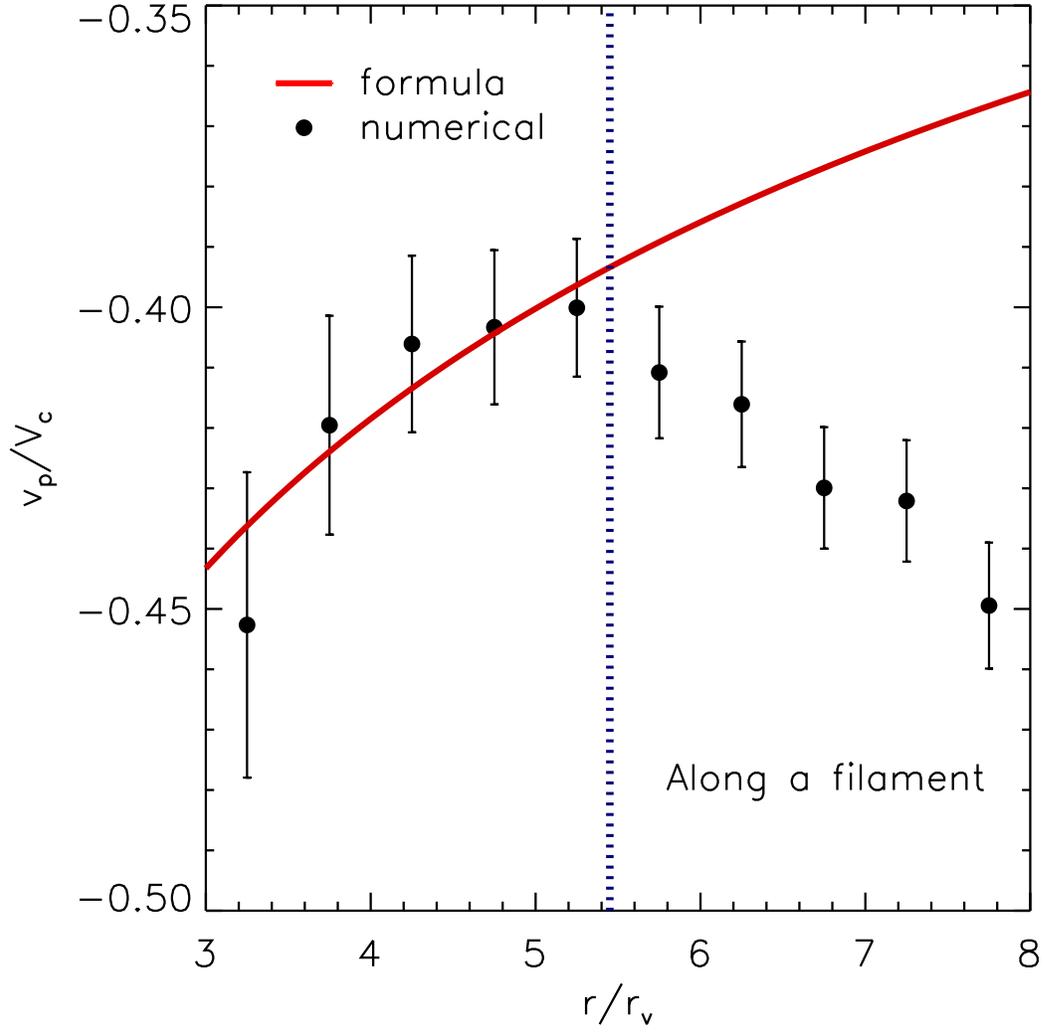}
\caption{Same as Figure \ref{fig:vpr_sheet} but for the case that the average peculiar velocity profile is 
constructed along the bound-zone filaments.}
\label{fig:vpr_fil}
\end{figure}

For the construction of $\tilde{v}_{\rm p}$ along the bound-zone filament, we also measure the azimuthal angles ($\phi$) of the 
position vectors, ${\bf r}$, of the bound-zone objects around each main group.
The bound-zone volume of each main group is first divided into $100$ conic subvolumes through simultaneously binning the values 
of $\theta$ and $\phi$. Counting the number of the bound-zone halos belonging to each subvolume, we locate the conic subvolume
which contains the maximum number of the bound-zone halos and designate it as the bound-zone filament.  
Using only those bound-zone halos belonging to the bound-zone filaments, we construct the average peculiar velocity profile, 
which is shown as black filled circles in Figure \ref{fig:vpr_fil}.  Note that the magnitude of $\tilde{v}_{\rm p}$ touches it minimum 
value at $\tilde{r}\sim 5.5$ and changes its trend, increasing with $\tilde{r}$ in the range of $\tilde{r}>5.5$. 
Equation (\ref{eqn:vpr}) fitted only to those numerical data points in the range of $\tilde{r}\le 5.5$ is shown as red solid line in 
Figure \ref{fig:vpr_fil}, where the two parameters are set at their best-fit values found to be 
$a=0.55\pm 0.01$ and $b=0.20\pm 0.01$. 
 
The distinct behaviors of $\tilde{v}_{\rm p}$ in the range of $\tilde{r}>5.5$ revealed by Figures \ref{fig:vpr_sheet} and \ref{fig:vpr_fil} 
clearly demonstrate that Equation (\ref{eqn:vpr}) with constant values of $a$ and $b$ fails to describe the peculiar velocity profiles 
constructed along the bound-zone sheets and filaments. Our explanation for this result is as follows. 
In the bound-zone around a main group, the halos would be affected not only by the gravity of the main group but also by 
that of the neighbor groups located outside the bound zone. When the peculiar velocity profile is constructed by taking the average 
over all directions, the gravitational forces from the neighbor groups would be cancelled out and thus have no net effect on the behaviors 
of the bound-zone peculiar velocity profile. However, when the construction of the peculiar velocity profile is made along the sheets 
and filaments, the gravitational forces from the neighbor halos no longer average out to zero but have non-zero net effect of 
flattening or increasing the magnitude of the peculiar velocities of the bound-zone halos at the distances larger than some 
threshold distance.  

The comparison between the two results in Figures \ref{fig:vpr_sheet} and \ref{fig:vpr_fil} indicates that when the bound-zone peculiar 
velocity profile is constructed along one dimensional filament, the profile in the range of $\tilde{r}>5.5$ becomes more severely 
contaminated by the gravitational influences of the neighbor groups outside the bound zone.  This result is in line with the recent finding 
of \citet{svensmark-etal15} that the gravitational forces of the surrounding large-scale structures quantified by the 
two halo terms in the density correlation function \citep{MF00,HW08} have the strongest effect on the peculiar velocities of the galaxies 
that fall into the clusters along the filaments.

\begin{figure}[tb]
\includegraphics[scale=1.0]{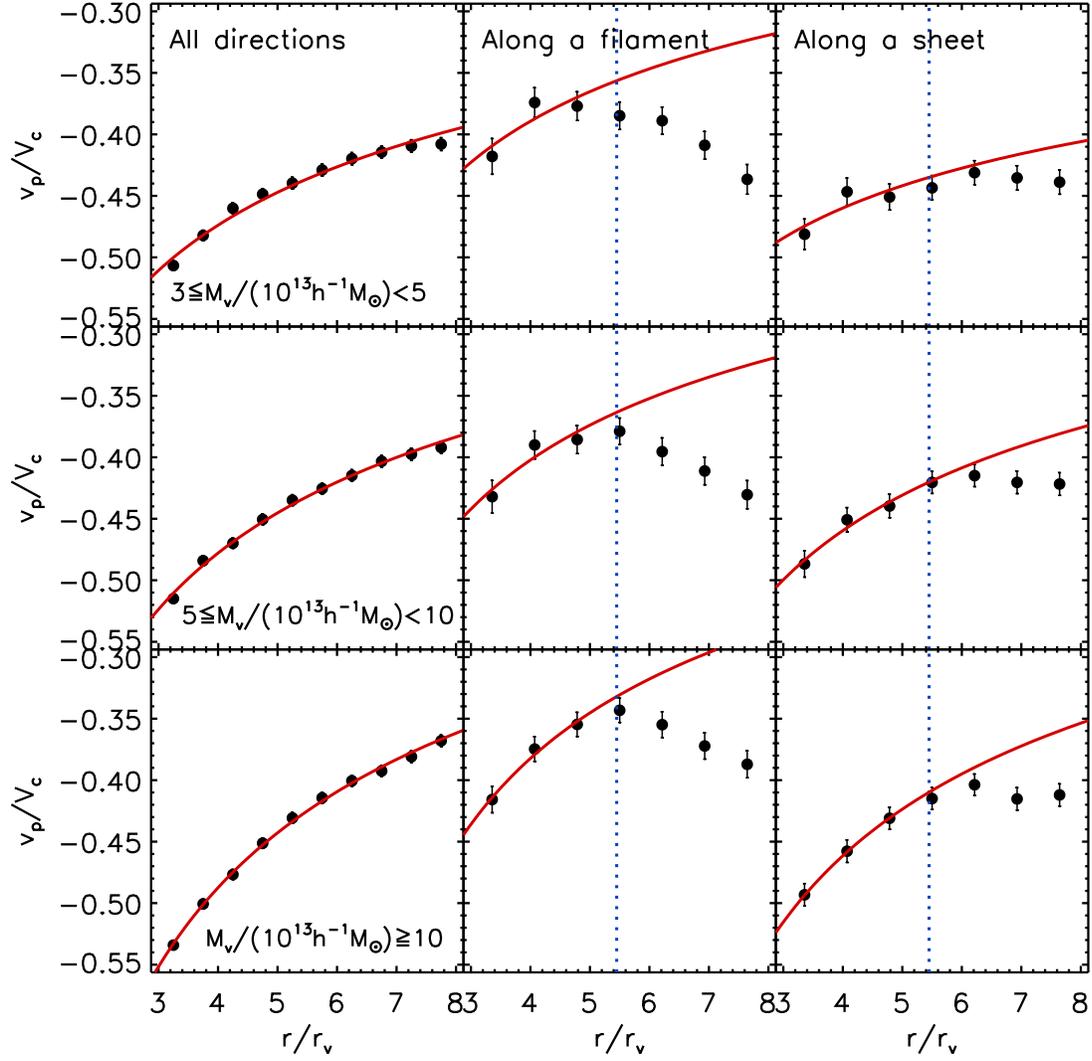}
\caption{Dependence of the average peculiar velocity profiles on the mass range.}
\label{fig:vpr_mweb}
\end{figure}

To see the trend of $\tilde{v}_{\rm p}$ with the variation of the range of $M_{\rm v}$, we also divide the values of $M_{\rm v}$ 
into three intervals and reconstruct $\tilde{v}_{\rm p}$ by using only those main groups whose masses belong to each 
$M_{\rm v}$-interval.  Figure \ref{fig:vpr_mweb} shows how differently the peculiar velocity profiles constructed along all directions 
and along the cosmic web behave for three cases of the $M_{\rm v}$ bins. The best-fit values of $a$ and $b$ as well as the numbers 
of the main groups for each case are listed in Table \ref{tab:all}.  As can be seen, the best-fit amplitude and slope  decrease as the mass 
scales of the main groups decrease \citep[see, also][]{cuesta-etal08}. 
For the case of the most massive groups with $M_{\rm v}\ge 10^{14}\,h^{-1}M_{\odot}$, the amplitude and slope of the 
bound-zone peculiar velocity profile constructed along all directions have the best-fit values of $a=0.88\pm 0.02$ and $b=0.43\pm 0.01$, 
which agree well with the original values determined by \citet{falco-etal14}. 
Whereas, the other two cases of the lower-mass groups yield substantially lower slopes and amplitudes than the original values, 
which implies that the lower masses are less central since the energy consideration requires the slope to be close to $b\sim 0.5$ 
for the central masses.

Figure \ref{fig:vpr_mweb} also displays in the middle and left panels how the shapes of the peculiar velocity profiles 
constructed along the cosmic web differ between the $M_{\rm v}$-bins and Table \ref{tab:all} lists the best-fit values of $a$ and $b$ 
for each case that the construction of $\tilde{v}_{\rm p}$ are made along the cosmic web. 
Note that the less massive a main group is, the more rapidly the magnitude of its bound-zone peculiar velocity profile constructed along 
filaments increases with distance in the section of $\tilde{r}> 5.5$.  The same logic provided in the above can also explain this result:  
The gravitational forces of the neighbor groups outside the bound zone should have stronger influences on the bound zones around the 
lower mass groups.

It is worth noting that the threshold value of the rescaled distances, say $\tilde{r}_{\rm c}$, at which the monotonic decreases of
the magnitudes of the peculiar velocity profiles constructed along the cosmic web come to a halt seem to be 
constant $r_{\rm c}\approx 5.5$. The threshold value of $\tilde{r}_{\rm c}$ represents the maximum rescaled distance up to which 
the gravitational effect of the neighbor groups outside the bound zone on $\tilde{v}_{\rm p}$ is negligible.  We make a serendipitous 
discovery that the threshold rescaled distance, $\tilde{r}_{c}$, coincides with the value of the following ratio:
\begin{eqnarray} 
\label{eqn:rsc1}
\frac{r_{\rm t, u}}{r_{\rm v}} &=& 
 \left(\frac{ 3M_{\rm v}G}{\Lambda c^{2}}\right)^{1/3}\times \frac{1}{r_{\rm v}} \\ 
 \label{eqn:rsc2}
&=& \left[\frac{4\pi\,\rho_{\rm c}\,\Delta_{\rm v}\,G}{\Lambda c^{2}}\right]^{1/3} \approx 5.43 
\end{eqnarray}
Here, we employ the relation of $M_{\rm v} = (4\pi/3)\Delta_{\rm v}\rho_{\rm c}r^{3}_{\rm v}$ where 
$\rho_{\rm c}$ is the critical density at the present epoch and $\Delta_{\rm v}\sim 100$ is the ratio of the mass density inside 
$r_{\rm v}$ to $\rho_{c}$ \citep{falco-etal14}.  
The dotted blue lines shown in Figures \ref{fig:vpr_sheet}-\ref{fig:vpr_mweb} correspond to this ratio, 
$r_{\rm t, u}/r_{\rm v}$.  

It may be worth discussing here why we did not follow the conventional scheme to identify the sheets and filaments in 
the bound zones around the main groups.  The Lagrangian perturbation theory describes that the sheets (filaments) form 
when the gravitational collapse proceeds anisotropically along the major (major and intermediate) principal axes of the 
deformation tensors calculated as the second derivaties of the perturvation potentials \citep{zel70}.  
Accordingly, the sheets (filaments) have been conventionally identified as the regions where the local tidal shear tensors have 
one (two) positive and two (one) negative eigenvalues \citep{hahn-etal07}.  

If we identified the sheets and filaments by using the conventional scheme, however, Equation (\ref{eqn:2dvpr}) would not be 
applicable since the success of Equation (\ref{eqn:2dvpr}) requires all of the bound-zone objects belonging to a sheet (or a filament) 
to possess the same inclination angles or at least to have the angles constrained in the small interval $[\cos\beta, \cos\beta+d\cos\beta]$. 
In other words, Equation (\ref{eqn:2dvpr}) is applicable only for the case that a bound-zone sheet (filament) possesses a flat plane-like 
(straight line-like) shape. Our scheme is deliberately devised to ensure that the identified sheets and filament have the required 
shapes. 

\section{TURN-AROUND RADII OF THE ISOLATED GROUPS}\label{sec:iso}

In this Section, we are going to estimate the turn-around radii of the main groups with the help of the methodology of 
\citet{lee-etal15} based on Equation (\ref{eqn:rt}). But, before embarking on the estimates of the turn-around radii, let us appreciate 
the implication of the results obtained in Section \ref{sec:web} that the bound-zone peculiar velocity profiles constructed along the 
cosmic web can be contaminated by the gravitational influence of its neighbor groups with comparable masses located outside the 
bound zone. 
The bottom line of Section \ref{sec:web} is that Equations (\ref{eqn:vpr}) and (\ref{eqn:2dvpr}) fail to describe the bound-zone 
peculiar velocity profiles constructed along the cosmic web. Given that the validity of Equations (\ref{eqn:vpr}) and (\ref{eqn:2dvpr}) 
is a key to the practical success of the methodology of \citet{lee-etal15}, it should be necessary to sort out those main groups whose 
peculiar velocity profiles are not severely contaminated by the gravitational field outside the bound zone. 
 
\begin{figure}[tb]
\includegraphics[scale=1.0]{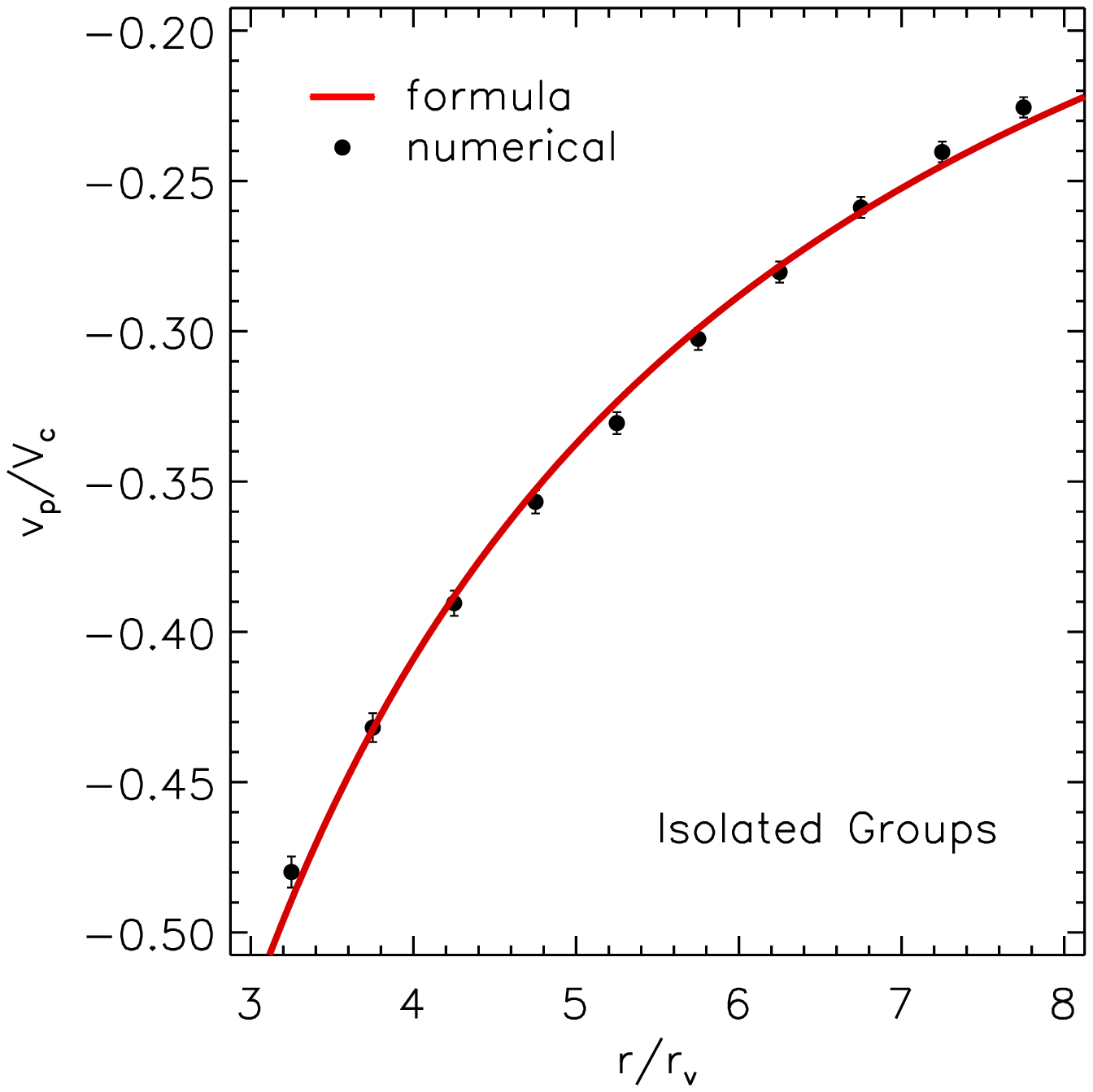}
\caption{Same as Figure \ref{fig:vpr_z0} but for the case of the isolated groups.}
\label{fig:vpr_iso}
\end{figure}

We suspect that for the case of {\it isolated} groups which have no neighbor groups with comparable masses in the regions near to 
the bound zone, the contamination of the peculiar velocity profiles due to the gravitational field outside the bound-zone may be 
attenuated to the negligible level. 
For each selected main group with $M_{\rm v}\ge 3\times 10^{13}\,h^{-1}M_{\odot}$, we search for its nearest neighbor 
group and measured the distance, $d_{s}$, between them. Then, those groups which satisfy $d_{s}\ge  d_{c}=15r_{\rm v}$ are 
identified as the isolated main groups.  Following the same procedures described in Section \ref{sec:web}, we construct 
the bound-zone peculiar velocity profiles averaged over a total of $1540$ isolated main groups. Figure \ref{fig:vpr_iso} shows the 
same as Figure \ref{fig:vpr_z0} but for the case of the isolated main groups. The best-fit parameters for this case are found to be 
$a=1.35\pm 0.03$ and $b=0.86\pm 0.02$. As can be seen,  the magnitude of the bound-zone peculiar velocity profile 
averaged over the isolated main groups decreases much more rapidly with $\tilde{r}$ than that averaged over all main 
groups shown in Figure \ref{fig:vpr_z0}.

\begin{figure}[tb]
\includegraphics[scale=1.0]{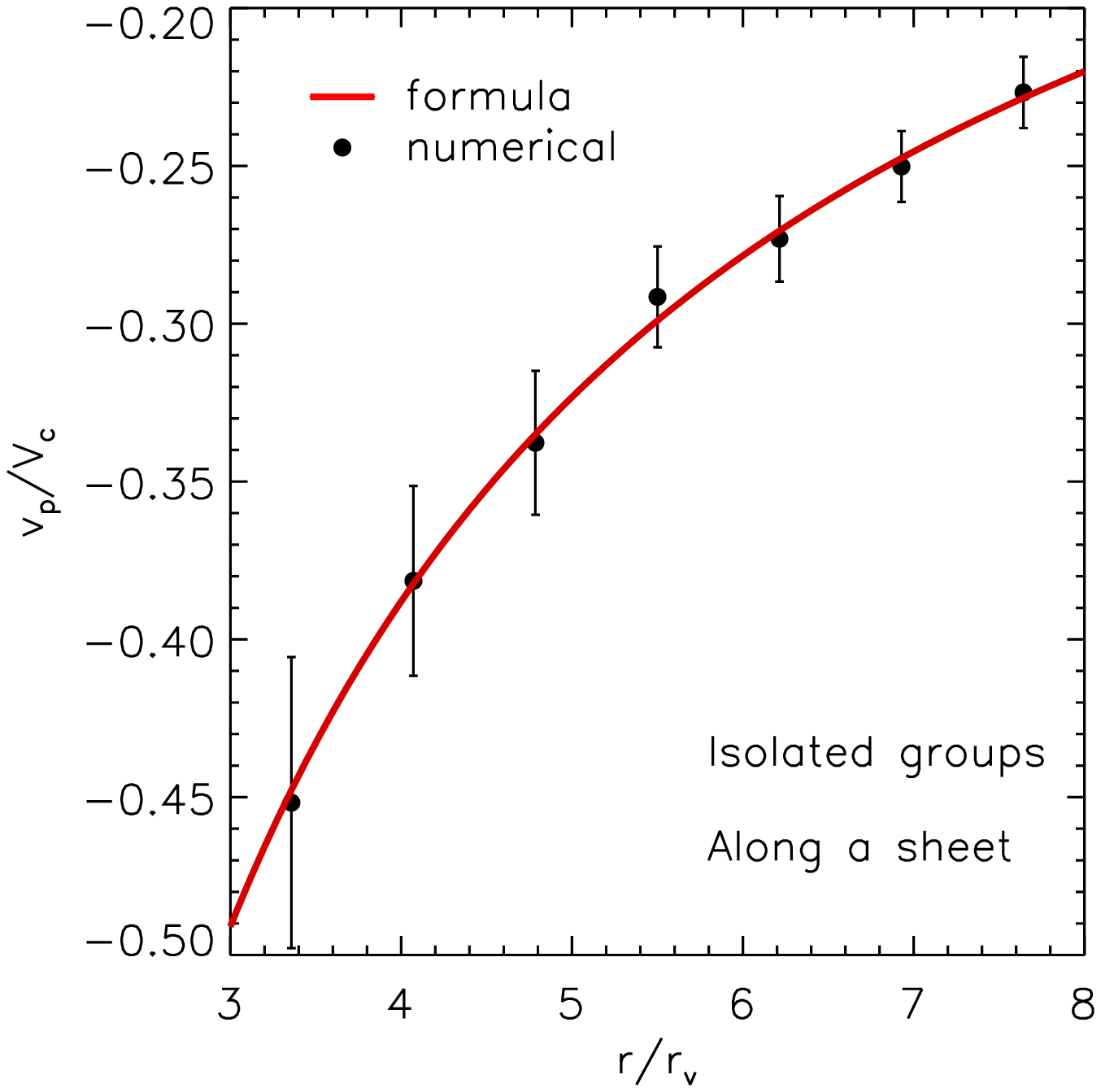}
\caption{Same as Figure \ref{fig:vpr_sheet} but for the case of the isolated groups.}
\label{fig:vpr_sheetiso}
\end{figure}

\begin{figure}[tb]
\includegraphics[scale=1.0]{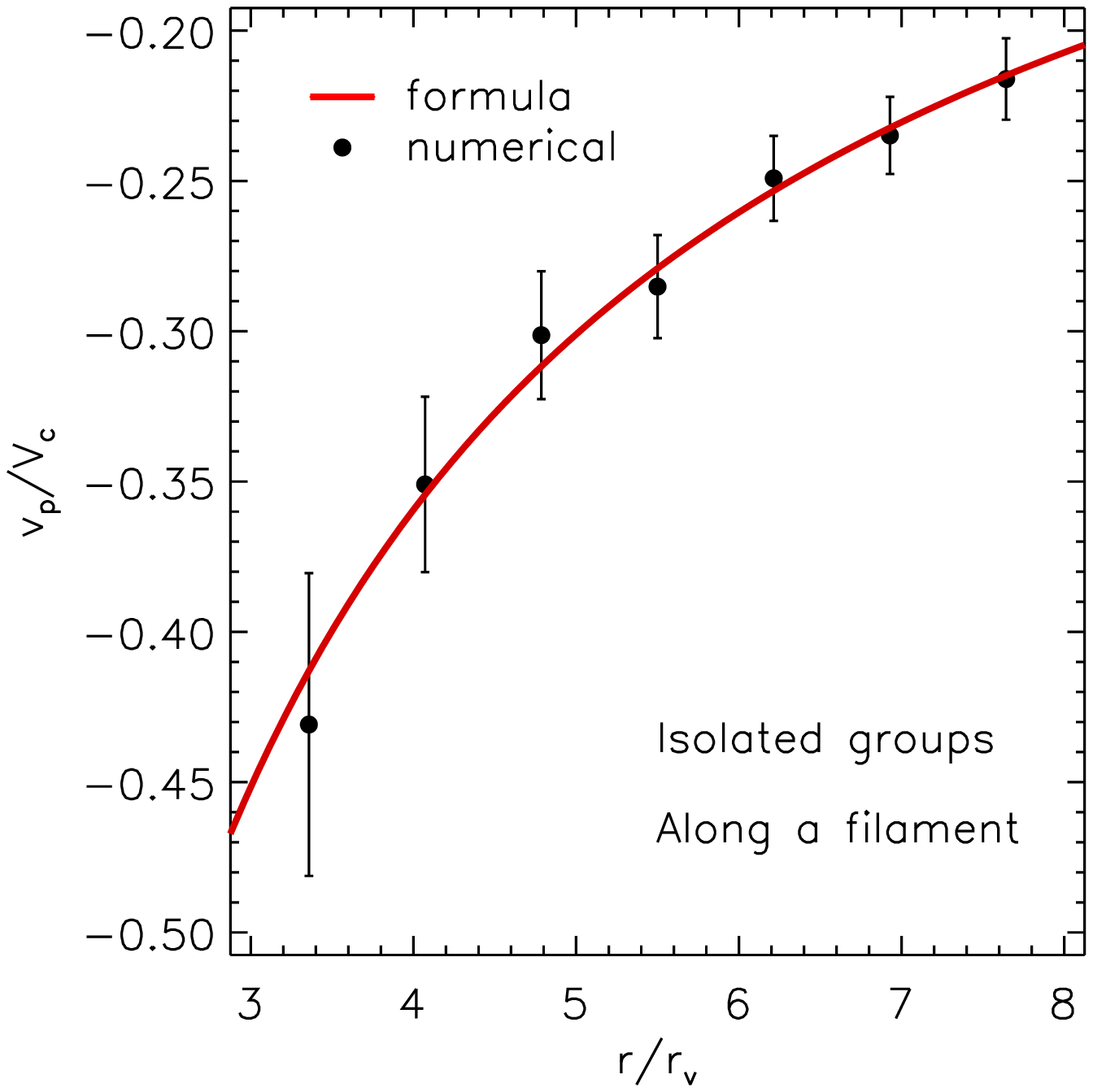}
\caption{Same as Figure \ref{fig:vpr_fil} but for the case of the isolated groups.}
\label{fig:vpr_filiso}
\end{figure}

Figures \ref{fig:vpr_sheetiso} and \ref{fig:vpr_filiso} show the same as Figures \ref{fig:vpr_sheet} and \ref{fig:vpr_fil} but 
for the cases of the isolated main groups, respectively.  The best-fit values are found to be 
$a=1.21\pm 0.22$ and $b=0.82\pm 0.10$ ($a=1.08\pm 0.38$ and $b=0.82\pm 0.21$) for the case of $\tilde{v}_{\rm p}(\tilde{r})$ 
constructed along the bound-zone sheets (filaments) around the isolated main groups. 
It is interesting to note that for the case of the isolated main groups, the magnitudes of $\tilde{v}_{\rm p}$ constructed 
along the cosmic web decrease monotonically in the whole bound zone range $3\le r/r_{\rm v}\le 8$ just as 
$\tilde{v}_{\rm p}$ constructed via isotropic average. This result indicates that the contamination caused by the gravitational 
influences of the neighbor groups on the bound-zone peculiar velocity profiles vanishes for the cases of the isolated main 
groups, as we have anticipated. We have tested the robustness of this result against the variation of $d_{c}$ and  
that only when $d_{c}\ge 15r_{\rm v}$, the profile $\vert\tilde{v}_{\rm p}\vert$ shows monotonic decrease with 
$\tilde{r}$ in the whole range of $\tilde{r}$.

\begin{figure}[tb]
\includegraphics[scale=1.0]{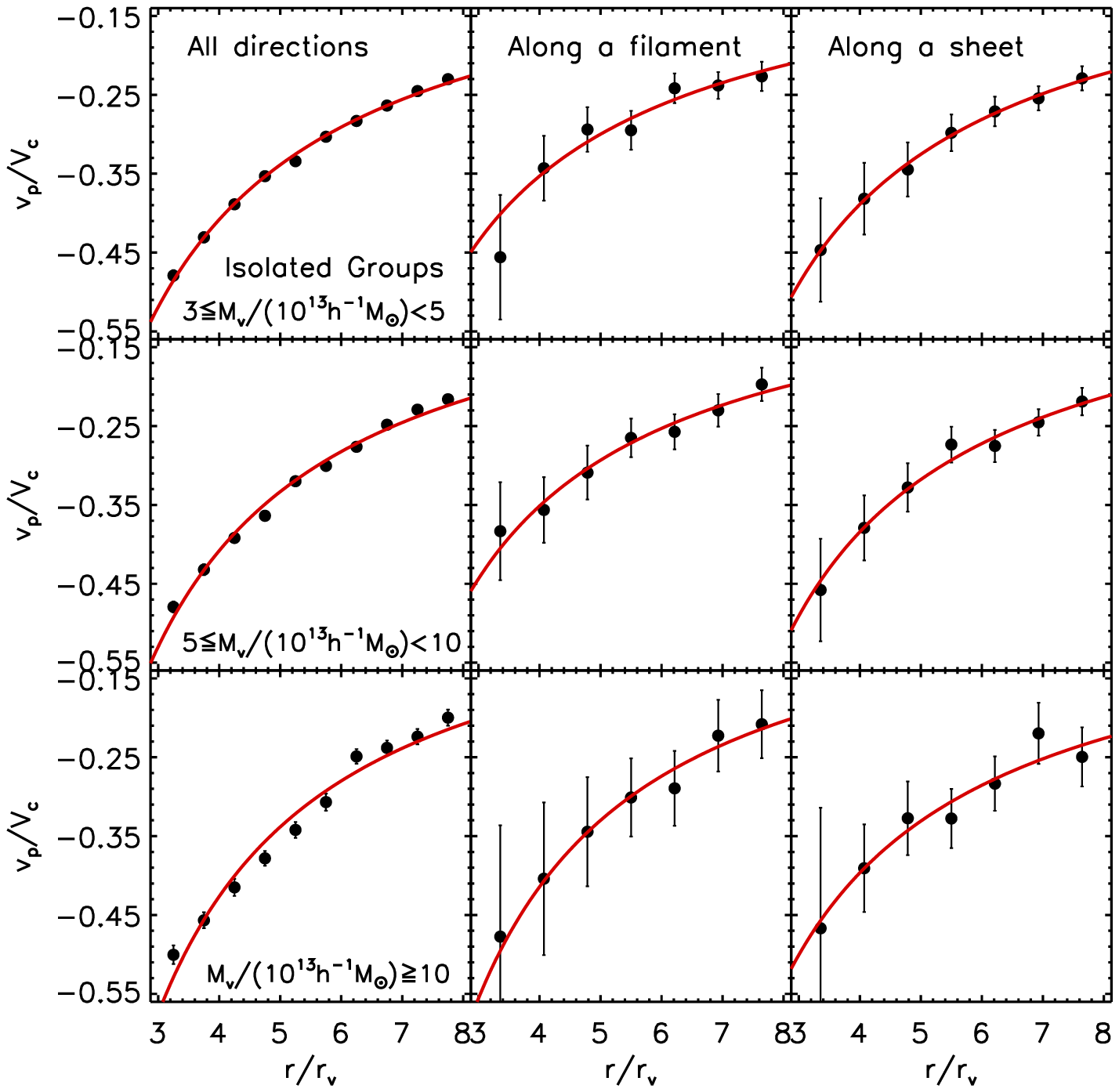}
\caption{Same as Figure \ref{fig:vpr_mweb} but for the case of the isolated groups.}
\label{fig:vpr_mwebiso}
\end{figure}

Figure \ref{fig:vpr_mwebiso} and Table \ref{tab:iso} show the same as Figure \ref{fig:vpr_mweb} and Table \ref{tab:all} but for the 
case of the isolated main groups, respectively.  It is important to note that for the cases of the isolated main groups the bound-zone 
peculiar velocity profiles do not show strong dependence on the mass scale $M_{\rm v}$, although the best-fit values of $a$ and 
$b$ suffer from large errors due to the small-number statistics (especially in the large-$M_{\rm v}$ bin).   
The results shown in Figures \ref{fig:vpr_iso}-\ref{fig:vpr_mwebiso} indicate that the analytic formula, Equation (\ref{eqn:vpr}), with 
fixed values of $a$ and $b$ can reliably describe the peculiar velocity profile in the whole bound zone range only when the 
condition of isolation is imposed on the selection of the main groups. 

Now, we are ready to estimate the turn-around radii of the isolated main groups by following the algorithm of \citet{lee-etal15}. 
First, we construct the peculiar velocity profile of  each individual isolated main group and determine the best-fit values of 
$a$ and $b$ by fitting Equation (\ref{eqn:vpr}) to each individually constructed peculiar velocity profile.  Putting the best-values 
of $a$ and $b$ into Equation (\ref{eqn:rt}), we estimate the turn radius $r_{\rm t}$ of each individual isolated main group. 
Then, we divide the range of $M_{\rm v}$ into $15$ bins and take the ensemble average of $r_{\rm t}$ over the isolated main 
groups whose virial masses belong to each $M_{\rm v}$-bin.  The one standard deviation scatter around the average is also 
calculated at each $M_{\rm v}$-bin as $\sigma = \left(N^{-1}_{\rm iso}\sum_{\alpha}\Delta r^{2}_{{\rm t},\alpha}\right)^{1/2}$ where 
$N_{\rm iso}$ is the number of the isolated main groups belonging to each $M_{\rm v}$-bin and $\Delta r_{{\rm t},\alpha}$ 
represents the difference of the turn-around radius of the $\alpha$-th main group from the ensemble average.  
The errors associated with the calculation of the ensemble average is also computed as $\sigma/\sqrt{N_{\rm iso}-1}$.
 
\begin{figure}[tb]
\includegraphics[scale=1.0]{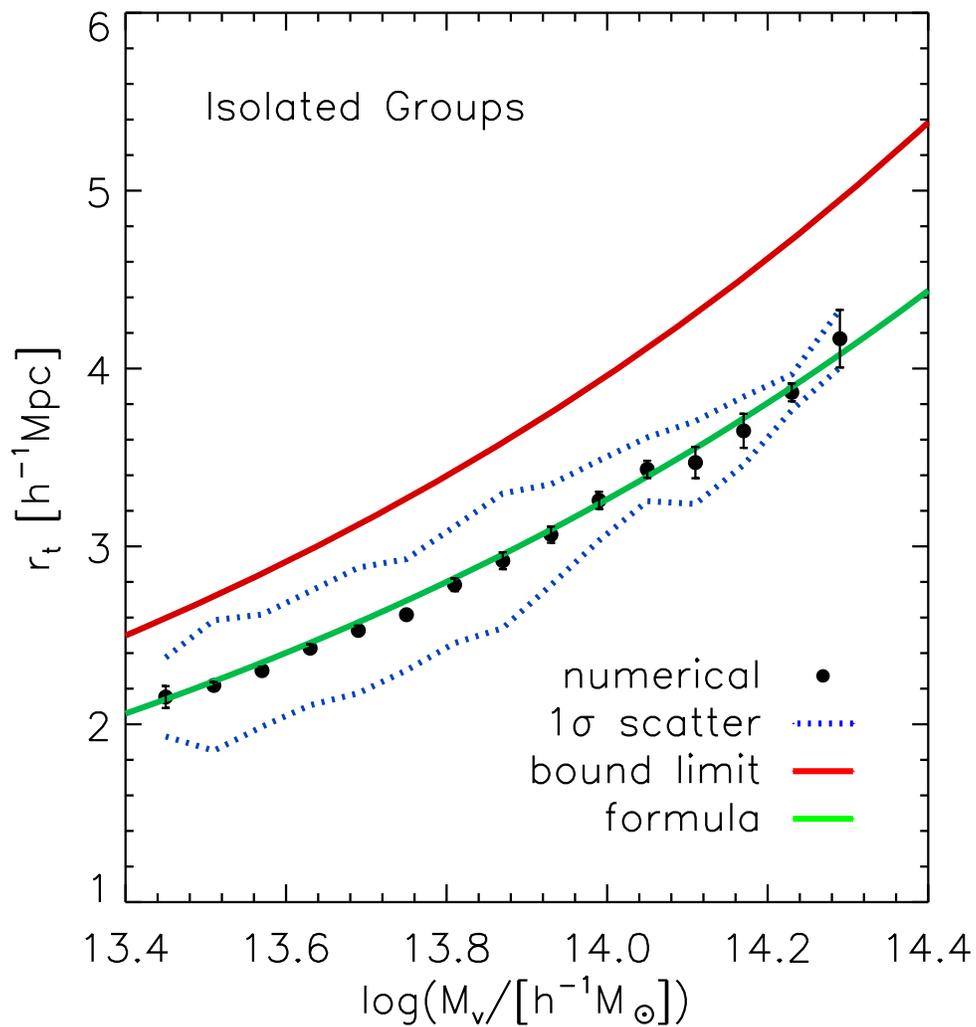}
\caption{Average turn-around radius as a function of the virial mass of the isolated groups. 
The black filled circles represent the mean value of the turn-around radius averaged over 
the central groups at each mass bin, the dotted blue line represents the one standard deviation 
scatter around the average value, the solid green line represent the original formula 
derived by, and the red solid line represents the bound-limit predicted by the Planck cosmology.}
\label{fig:ta_iso}
\end{figure}

The turn-around radii $r_{\rm t}$, estimated by the algorithm of \citet{lee-etal15} is plotted versus the virial masses 
$M_{\rm v}$ of the isolated main groups as the black filled circles with errors in Figure \ref{fig:ta_iso}. The red solid line is the 
spherical bound limit, $r_{\rm t, u}$, given in Equation (\ref{eqn:rt_u}) predicted by the Planck cosmology. The blue dotted line is the 
one standard deviation scatter $\sigma$ around the average values of $r_{\rm t}$. This result clearly reveals that the spherical 
bound limit (red solid line) is much higher than the estimated turn-around radii in the whole mass range, implying that a bound-
violation (i.e., an event of $r_{\rm t}\ge r_{\rm t, u}$) seldom occurs in the Planck universe. Note that a bound violation is relatively 
less rare on the low mass scale ($M_{\rm v}\le 5\times 10^{13}\,h^{-1}M_{\odot}$), which can be attributed to the fact that the 
spherical symmetry of the gravitational collapse process breaks down more severely on the low mass scales \citep{bernardeau94}. 

The green solid line in Figure \ref{fig:ta_iso} represents the theoretical turn-around radius as a function of $M_{\rm v}$ 
evaluated by solving Equation (\ref{eqn:rt}) whose two parameters are set at the best-fit values of $a=1.35$ and 
$b=0.86$ determined from the average (not individual) peculiar velocity profile (see Figure \ref{fig:ta_iso}).  
While the black filled circles with errors have been obtained by applying the method of \citet{lee-etal15} to the bound-zone 
peculiar velocity profiles of the individual isolated groups, the green solid line has been theoretically evaluated by 
applying the method of \citet{lee-etal15} to the mean bound-zone peculiar velocity profile averaged over all of the isolated 
groups. The good agreement between the black filled circles and the green solid line indicates  implies the robustness and solidity 
of the algorithm of \citet{lee-etal15} for the estimation of the turn-around radii.
 
\begin{figure}[tb]
\includegraphics[scale=1.0]{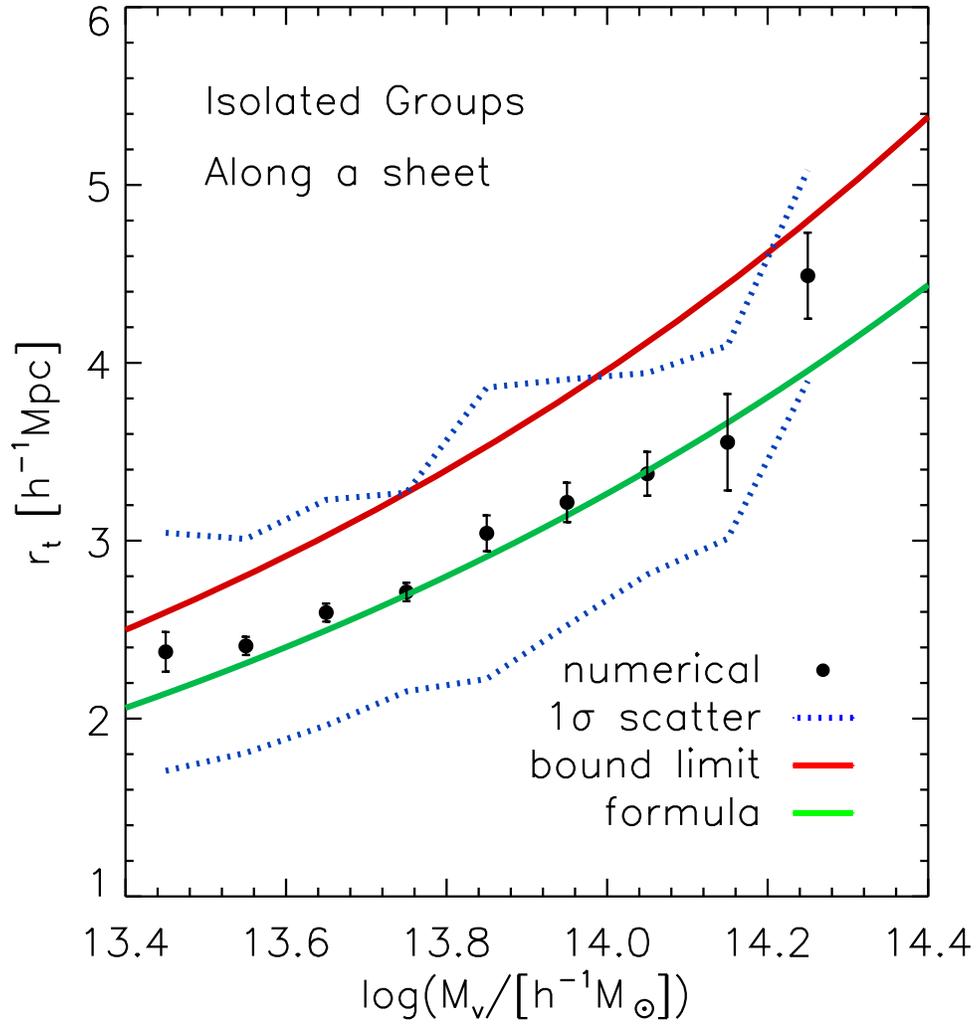}
\caption{Same as Figure \ref{fig:ta_iso} but for the case of the bound-zone sheets.}
\label{fig:ta_sheetiso}
\end{figure}

\begin{figure}[tb]
\includegraphics[scale=1.0]{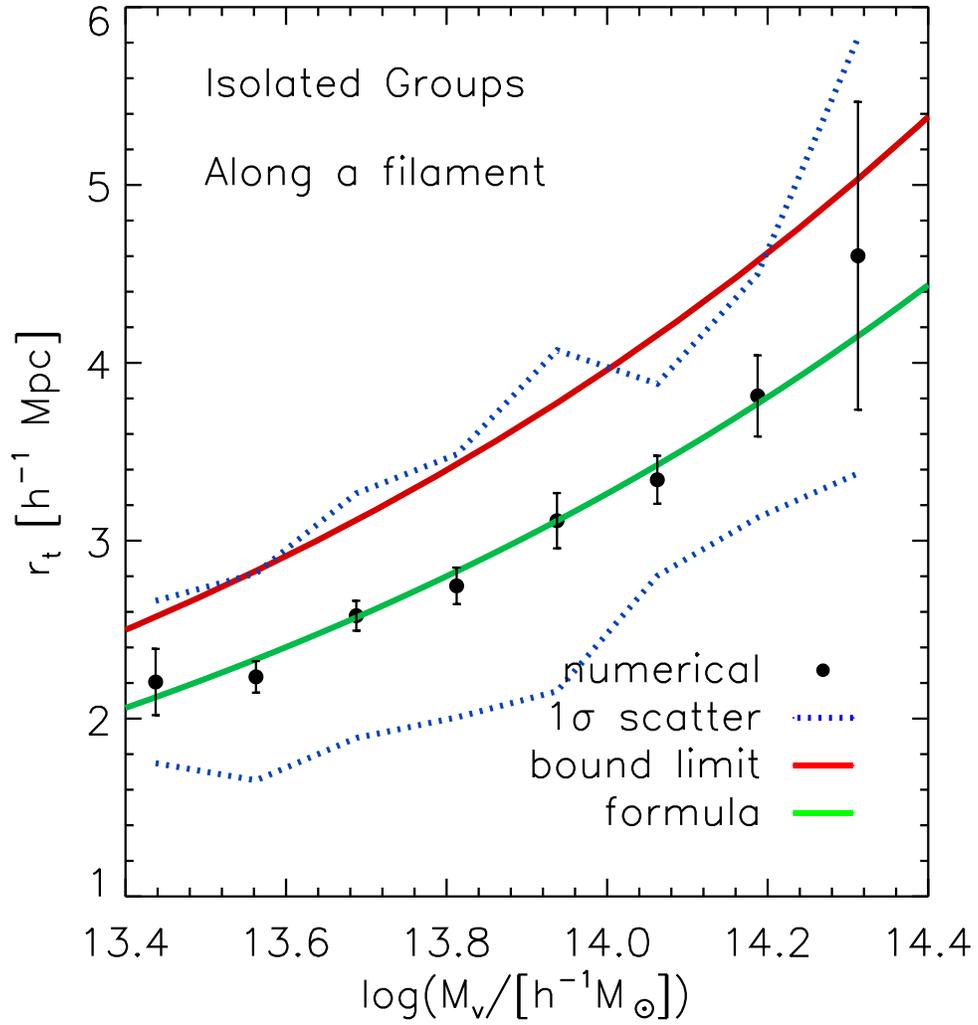}
\caption{Same as Figure \ref{fig:ta_iso} but for the case of the bound-zone filaments.}
\label{fig:ta_filiso}
\end{figure}

Now, we would like to estimate $r_{\rm t}$ of the isolated main groups along the cosmic web. 
Basically, we repeat the same procedure for the estimation of $r_{\rm t}(M_{\rm v})$ as described in the above but 
by using the individual peculiar velocity profiles $\tilde{v}_{\rm p}$ constructed along the sheets and the filaments.  
Figures \ref{fig:ta_sheetiso} and \ref{fig:ta_filiso} show the same as Figure \ref{fig:ta_iso} but for the case that the turn-around radii 
of the individual isolated main groups are estimated along the sheets and the filaments, respectively.  To plot the green solid 
line in Figure \ref{fig:ta_sheetiso} (Figure \ref{fig:ta_filiso}) the method of \citet{lee-etal15} has been applied to the mean peculiar 
velocity profile shown in Figure \ref{fig:vpr_sheetiso} (Figure \ref{fig:vpr_filiso}). As can be seen, the ensemble averages of the 
turn-around radii estimated along the cosmic web are still lower than the spherical upper limit in the whole mass range. 

The comparison of Figures \ref{fig:ta_sheetiso} and \ref{fig:ta_filiso} with Figure \ref{fig:ta_iso} reveals that the values of 
$r_{\rm t}$ estimated along the cosmic web vary in a wider range and that the spherical bound limit (red solid line) is placed 
within one standard deviation scatter from the ensemble average. In other words, if the turn-around radii of the individual 
isolated main groups are estimated along the cosmic web, it will be less rare to witness the bound violations even in a 
$\Lambda$CDM universe.  Hence, an observation of a single bound violation of an individual object would not challenge 
the standard model. Only if the spherical bound limit is found to be violated by the {\it average} value of the 
turn-around radius rather than by the individual ones, it should be regarded as an anomaly. 

\section{SUMMARY AND DISCUSSION}\label{sec:con}

We have numerically estimated the turn-around radii $r_{\rm t}$ of the massive groups with virial mass 
$M_{\rm v}\ge 3\times 10^{13}h^{-1}M_{\odot}$ identified in the SMDPL \citep{mdark_sim,rockstar_er} by constructing the peculiar 
velocity profiles of the bound-zone halos with $M_{\rm v}\ge 5\times 10^{10}\,h^{-1}M_{\odot}$ around the groups. 
The shapes and behaviors of the peculiar velocity profiles have been found to vary with the directions along which the 
constructions were made.  When the bound-zone halos distributed along all directions are used for the constructions, 
the magnitudes of the resulting peculiar velocity profiles monotonically decrease with distance, being well approximated 
by the simple formula of \citet{falco-etal14} with two characteristic parameters. 

However, when only those bound-zone halos distributed along some confined directions around the groups are used for the 
constructions, the resulting peculiar velocity profiles exhibit distinct features: Instead of monotonically decreasing with distance, 
the magnitudes of the profiles reach the minimum values at some critical distances $r_{\rm c}$ and remain constant (show increment) 
with distance after $r_{\rm c}$ when the constructions of the profiles are made from those bound-zone halos distributed along the sheets 
(filaments). 
Explaining that this distinctly unique behaviors of the bound-zone peculiar velocity profiles constructed along the sheets and the 
filaments should originate from the external gravitational effect of the neighbor massive groups, we have imposed the condition 
of isolation on the selection of the main groups to avoid any contamination in the estimates of their turn-around radii due to the 
external gravity.  As anticipated, it has been shown that the bound-zone peculiar velocity profiles around the isolated main groups 
exhibit monotonic decrement with distance, even when the profiles are constructed along the sheets and the filaments. 
 
Finally, the average turn-around radii of the isolated main groups have been estimated from the bound-zone peculiar velocity 
profiles with the help of the algorithm developed by \citet{lee-etal15} and found to be well below the spherical bound limit 
predicted by the $\Lambda$CDM cosmology in the whole mass range.   We have also quantitatively shown that it is quite rare but 
not forbidden for individual groups to violate the bound limit especially on the relatively low-mass scale 
$\le  5\times10^{13}\,h^{-1}M_{\odot}$ and 
that the rarity of the bound-violations diminishes when the turn-around radii are estimated along the cosmic web.  
Given our result, it has been suspected that the bound-violation of NGC 5353/4 group reported by \citet{lee-etal15} may 
not be interpreted as a counter-evidence against the $\Lambda$CDM model at the moment.  

It is worth discussing our serendipitous discovery of the coincidence between $r_{\rm c}$ and $r_{\rm t, u}$ for the case of 
the non-isolated main groups. When the bound-zone peculiar velocity profiles are constructed along the filaments around the 
{\it non-isolated} main groups, the threshold distances, $r_{\rm c}$, at which the magnitudes of the bound-zone peculiar velocities 
touch their minimum values coincide with the spherical upper limit, $r_{\rm t,u}$, at all mass scales. Given that the spherical 
upper limit, $r_{\rm t,u}$ is a direct indicator of the amount of $\Lambda$ as shown in Equation (\ref{eqn:rt_u}), our result hints 
at the possibility of using $r_{\rm c}/r_{\rm v}$ as another local test of the $\Lambda$CDM cosmology.  
Constructing the average bound-zone peculiar velocity profile along the filaments around the non-isolated massive groups from the 
direct observables with the help of Equation (\ref{eqn:2dvpr}) and locating where the magnitude of the average bound-zone peculiar 
velocity profile reaches the minimum value, we can determine $r_{\rm c}/r_{\rm v}$. Then, equating it to the theoretical constant 
given in Equation (\ref{eqn:rsc1}), we can estimate the amount of $\Lambda$. It would also allow us to constrain the equation of 
state of dark energy since the value of the spherical upper limit changes if the equation of state of dark energy deviates from $-1$ 
\citep{pavlidou-etal14}. To make a robust estimate of $r_{\rm c}$ from the average bound-zone peculiar velocity profile from 
observations, it will be definitely necessary to use a large sample of the massive groups observed in the local universe, which 
is the direction of our future work.  

\acknowledgments

We are very grateful to our anonymous referee for very useful comments which helped us improve the original manuscript. 
This work was initiated during the workshop on "Near Field Cosmology 2016" held at the Obergurgl University Centre in Tyrol, Austria from 
March 29 - April 3, 2016. We thank Stefan Gottl\"ober and the other organizers of the workshop for making stimulating discussions among 
the participants possible during the workshop. This study made use of data from the CosmoSim database (http://www.cosmosim.org/), 
hosted and maintained by the Leibniz-Institute for Astrophysics Potsdam (AIP).  The MultiDark-Planck (MDPL) and the BigMD simulation 
suite have been performed on the Supermuc supercomputer at LRZ using time granted by PRACE. 
J.L. acknowledges the support of the Basic Science Research Program through the NRF of Korea funded by the Ministry of Education 
(NO. 2013004372).  J.L. was also partially supported by a research grant from the National Research Foundation (NRF) of Korea to the 
Center for Galaxy Evolution Research (NO. 2010-0027910).
G.Y. thanks MINECO/FEDER for financial support under research grants AYA2015-63810-P and AYA2012-31101.

\clearpage
\clearpage
\clearpage
\clearpage
\clearpage
\clearpage
\clearpage
\clearpage
\clearpage
\clearpage
\clearpage
\clearpage
\clearpage
\begin{deluxetable}{ccccc}
\tablewidth{0pt}
\setlength{\tabcolsep}{5mm}
\tablecaption{Mass range ($M_{\rm v}$) and number ($N_{\rm g}$) of the main groups, the direction along which the 
bound-zone peculiar velocity profile is reconstructed, the best-fit parameters of the analytic formula, $a$ and $b$ in 
Equation (\ref{eqn:vpr}).}
\tablehead{$M_{\rm v}$ & $N_{\rm h}$ & Direction &  $a$ & $b$ \\ 
$(10^{13}h^{-1}\,M_{\odot})$& & & &}
\startdata
$[3, 5)$ &  $3975$& All & $0.68^{+0.01}_{-0.02}$ & $0.26^{+0.01}_{-0.02}$\\
$[3, 5)$  &  $3975$ & Sheet & $0.59\pm 0.09$ & $0.18\pm 0.10$\\
$[3, 5)$ & $3975$ & Filament & $0.58\pm 0.12$ & $0.29\pm 0.14$\\
\hline
$[5, 10)$ & $2842$ & All & $0.74^{+0.01}_{-0.02}$ & $0.32^{+0.01}_{-0.02}$\\
$[5, 10)$  & $2842$ & Sheet & $0.69\pm 0.09$ & $0.29\pm 0.09$\\
$[5, 10)$ & $2842$ & Filament & $0.63\pm 0.12$ & $0.33\pm 0.12$\\
\hline
$[10, \infty)$ & $1659$ & All & $0.88\pm 0.02$ & $0.43\pm 0.01$\\
$[10, \infty)$ & $1659$ & Sheet & $0.79\pm 0.09$ & $0.38\pm 0.08$\\
$[10, \infty)$& $1659$ & Filament & $0.72\pm 0.11$ & $0.46\pm 0.11$ \\
\enddata
\label{tab:all}
\end{deluxetable}
\clearpage
\begin{deluxetable}{ccccc}
\tablewidth{0pt}
\setlength{\tabcolsep}{5mm}
\tablecaption{Same as Table \ref{tab:all} but for the case of the isolated main groups.}
\tablehead{$M_{\rm v}$ & $N_{\rm h}$ & Direction &  $a$ & $b$ \\ 
$(10^{13}h^{-1}\,M_{\odot})$& & & &}
\startdata
$[3, 5)$ &  $1088$& All & $1.30\pm 0.04$ & $0.83\pm0.02$\\
$[3, 5)$  &  $1088$ & Sheet & $1.18\pm 0.34$ & $0.80\pm 0.14$\\
$[3, 5)$ & $1088$ & Filament & $0.97\pm 0.33$ & $0.73\pm 0.17$\\
\hline
$[5, 10)$ & $407$ & All & $1.44\pm 0.05$ & $0.91\pm0.02$\\
$[5, 10)$  & $407$ & Sheet & $1.25\pm 0.35$ & $0.85\pm 0.14$\\
$[5, 10)$ & $407$ & Filament & $1.10\pm 0.37$ & $0.81\pm 0.17$\\
\hline
$[10, \infty)$ & $45$ & All & $1.80\pm 0.10$ & $1.04\pm 0.04$\\
$[10, \infty)$ & $45$ & Sheet & $1.22\pm 0.09$ & $0.80\pm 0.30$\\
$[10, \infty)$& $45$ & Filament & $1.70\pm 0.16$ & $1.01\pm 0.35$ \\
\enddata
\label{tab:iso}
\end{deluxetable}
\end{document}